\documentclass[useAMS,usenatbib]{mn2e}
\usepackage{graphicx}
\usepackage{txfonts}
\usepackage{color}
\bibpunct{(}{)}{;}{a}{}{,}

\input epsf.sty

\usepackage{hyperref}
\usepackage{aas_macros}
\usepackage{natbib}
\bibpunct{(}{)}{;}{a}{}{,}

\title[Thermal instability in the mini-spiral region]{Conditions for the Thermal 
Instability in the Galactic Centre Mini-spiral region}

\author[A. R\'o\.za\'nska et al.]{A. R\'o\.za\'nska$\!^{1}$\thanks{E-mail:
agata@camk.edu.pl}, B. Czerny$^{1}$, D. Kunneriath$^{2}$, T. P. Adhikari$^{1}$, 
V. Karas$^{2}$, M. Mo\'scibrodzka$^{3}$  \\
$^{1}$ N. Copernicus Astronomical Center, Bartycka 18, 00-716 Warsaw, Poland\\
$^{2}$ Astronomical Institute, Academy of Sciences, Bo\v{c}n\'{\i} II 1401, CZ-14131 Prague, Czech Republic \\
$^{3}$ Department of Astrophysics/IMAPP,  Radboud University Nijmegen,P.O. Box 9010, 
6500 GL Nijmegen, The Netherlands  \\}
\begin{document}

\date{Accepted ..........}

\pagerange{\pageref{firstpage}--\pageref{lastpage}} \pubyear{2002}

\maketitle

\label{firstpage}

\begin{abstract}
We explore the conditions for the thermal instability to operate in the mini-spiral region
 of the Galactic centre (Sgr A*), where both the hot and cold media are known to coexist. 
The photoionisation Cloudy€™ calculations are performed for different physical states of plasma. 
We neglect the dynamics of the material and concentrate on the study 
of the parameter ranges where the thermal instability may operate, 
taking into account the past history of Sgr A* bolometric luminosity. 
We show that the thermal instability does not operate at the present very low level
of the Sgr A* activity.  However, Sgr A* was much more luminous in the past.
For the highest luminosity states the two-phase medium can be created up to 1.4 pc 
from the centre. The presence of dust grains tends to suppress the instability, 
but the dust is destroyed in the presence of strong radiation field and hot plasma. 
The clumpiness is thus induced in the high activity period, 
and the cooling/heating timescales are long enough to preserve later the 
past multi-phase structure. The instability enhances the clumpiness of the mini-spiral 
medium and creates a possibility of episodes of enhanced accretion of cold 
clumps towards Sgr A*. The mechanism determines the range of masses and sizes of clouds; 
under the conditions of Sgr A*, 
the likely values come out $1$-–$10^2M_{\oplus}$ for the cloud typical mass.
\end{abstract}


\section{Introduction}

The thermal instability can develop when an interstellar medium is irradiated by 
sufficiently intense external source. In a certain range of parameters multiple 
phases can arise and thermally different states can stay in thermal equilibrium \citep{field1965}. 
Near a supermassive black hole a two-phase medium forms spontaneously at certain 
range of the central luminosities and distances from the centre  \citep{krolik1981}.

The interstellar medium has a complex multiphase structure which is not yet
fully understood \citep[for review, see][]{cox2005}. 
This structure is important from the point of
view of feeding the Milky Way's central supermassive black hole as well 
as the possibility of an accompanying outflow. A similar mechanism is expected
to operate in the other low luminosity galactic nuclei.

Some aspects of this multiphase structure were
already studied by \citet{field1965} who considered the thermal equilibrium of the
radiatively heated and cooled plasma. He showed that for a certain range of
the parameters the plasma is thermally unstable, and the colder, denser matter
may coexist with the rarefied hot medium in pressure equilibrium due to
different cooling/heating mechanisms dominating at high and low gas
densities.  This theory was further developed in a number of papers,
where the effect of steady evaporation or condensation of the cold phase was taken into account
\citep{cowie77,mckee77,begelman83,begelman1990}.  The thermal instability and eventual 
cloud evaporation predominantly depend on the heating and cooling processes
 taken into account in the  energy equilibrium equation. In the case of illuminated accretion 
disks viscous heating should be considered \citep{rozanska96,rozanska99}.

The next major step was done in series of three papers 
\citep{barai11,barai12,moscibrodzka13}, 
where the authors combined the simple theory of
spherical (Bondi) accretion with the theory of radiative instability due to
preheating of the accretion flow by the X-ray flux generated by accretion
close to the black hole. In all three papers, the authors performed careful
numerical simulations of the dynamics of the accretion flow with the proper
treatment of heating/cooling mechanisms. These results showed that a two-phase medium
formed spontaneously at some ranges of the central luminosity/distance from the
black hole \citep{barai11}, the colder clumps formed filaments which accreted 
faster than the surrounding hot plasma \citep{barai12}. 
These filaments may break into smaller cloudlets, which may be revealed only if 
numerical method allows for the required resolution \citep{moscibrodzka13}.

The Milky Way interstellar medium is the best studied case of complex cosmic
environment, and the multi-phase
region in the close vicinity of the central massive black hole is clearly detected
\citep{zhao2009}.
The accretion onto Sgr~A* is very complex, not only due to the hot
Bondi-type flow but also likely due to occasional accretion events of colder
clumps. The argument for the second aspect comes both from the currently
observed approach of the G2 cloud and also from the observed X-ray echo from
the large molecular clouds surrounding Sgr~A* which allows to constrain much
higher activity of Galactic centre (GC) in the past 
\citep{sunyaev93,koyama96,ponti2010,capelli2012}. The time profile of this activity
level with rather large luminosity, of order of $10^{40}$ erg s$^{-1}$ for a few
hundred years followed by a rapid fast decay is not easy to model, and the
suggested explanation was through an accretion of a chain of similar clouds
\citep{czerny13a}. Thus the question appears whether the mechanism considered
by \citet{barai11,barai12} and \citet{moscibrodzka13} can indeed apply to
the central parts of Sgr~A* and drive this behaviour.

\citet{czerny2013b} emphasize that the amount of material contained in the 
mini-spiral is sufficient to sustain the past luminosity of Sgr~A* at the 
required level. The accretion episodes of relatively dense gas from 
the mini-spiral passing through a transient accretion ring at about $10^4$ 
gravitational radii of the supermassive black hole provide a viable scenario
for the bright phase of the Galactic center. What remains to clarify is an
important aspect of the mechanism of the angular momentum loss that must
operate in order to bring this material further down to the horizon.

The observations of the circumnuclear region of Sgr~A* show a reservoir of cold
clumps in the mini-spiral. This material has a rather high
angular momentum so the whole picture is more complex than the quasi-spherical
dynamical setup of the previous works. Therefore, in the present paper we
entirely neglect the dynamics of the material and concentrate on the simple
study of the parameter ranges where the thermal instability may operate, taking into
account the varying X-ray flux of Sgr~A* in the past few hundred years.
Our approach is complementary to numerical simulations  
\citep{alig2013}, where the authors argue that the mini-spiral itself formed 
in a collision of a molecular cloud with
the circumnuclear disk. We show how those clouds can form due to thermal instabilities 
in the strong radiation field. 

We use Cloudy\footnote{http://www.nublado.org/} photoionisation code 
\citep{ferland2013} to:
(i) set up a simplified model of the ISM in the GC to investigate the
conditions required to obtain accretion driven by thermal instabilities,
(ii) calculate instability curves for different luminosity states of Sgr~A*
to determine if its past high luminosity state inferred from X-ray
reflection clouds is sufficient to create a two-phase medium in the
mini-spiral region,
(iii) study the influence of dust grains in the mini-spiral region on the
creation of thermal instabilities.

We show that for the highest luminosity states, the two-phase medium can be created
up to 1.4 pc away from the GC.
Timescales of most clouds are of the order of 10$^{2-6}$ years, depending 
on the cloud density. The extent of the two-phase medium depends on the 
value of gas pressure of the hot phase plasma. Adopting the gas pressure, of a 
typical Bondi flow, we calculated instability 
strips for several highest-luminosity states of Sgr~A*. 
Finally, we demonstrate that the dust content does not allow the emergence of 
thermal instability, but 
on the other side, dust should quickly evaporate in such hot Bondi 
flow \citep{draine1979}.

The structure of the paper is as follows: in Sec.~\ref{sec:obs} 
we summarize the morphology of the mini-spiral known from current 
observations, in Sec.~\ref{sec:met} we describe the method of
photoionisation calculations. 
All results are presented in Sec.~\ref{sec:res},~\ref{sec:two}, 
and discussed in Sec.~\ref{sec:dis}. 
 Final conclusions are given in Sec.~\ref{sec:concl}.

\section{The material content of the mini-spiral}
\label{sec:obs}

The interstellar medium in the central 10~pc of the GC is a complex 
mixture of ionized, neutral and molecular gas, and dust that is heated up 
by its interaction with the central supermassive black hole, Sgr~A*, and also directly by the 
stars in the 
Nuclear Stellar Cluster (NSC) in the region. 
The circumnuclear disc is a ring of $\simeq10^5M_\odot$
of neutral atomic and molecular material 
\citep[see][and references therein]{vollmer2002,chris2005}
 extending from about 1.5-10 pc, 
which encompasses Sgr~A~West, the thermal HII emission region, the brightest part 
of which is the so-called mini-spiral. The mini-spiral consists of three streams of ionized 
gas (the Northern Arm, Eastern Arm and the Western Arc), with most of the gas predominantly 
in a Keplerian orbit with observed radial velocity components spanning from $+200$ to $-415$ km/s, 
with a few significant deviations. The Northern and Eastern arms
 appear to collide about 0.1--0.2 pc behind Sgr~A* \citep{zhao2010}. 
The partially ionized gas in the mini-spiral arms have number densities 
ranging $3$--$21 \times10^4$ cm$^{-3}$, with temperatures in the interval 5000--13\,000~K 
\citep{zhao2009}. The mini-spiral also contains dust with temperatures in the range of  
200--300 K \citep{cotera1999}. The fraction of material in the form of dust is 
low ($\sim 0.50 M_{\odot}$) for the central $1.5\times1.5$
parsec region \citep{gezari85} when compared to the $\sim 160 M_{\odot}$ ionized gas 
in the central $3\times3$ pc region \citep{liszt2003}.

From the {\it Chandra} data \citep{baganof2003}, we have an estimate of local
density and temperature of the hot, fully ionized plasma: direct map analysis
implies $n_{\rm e} = 130$ cm$^{-3}$, $T_{\rm e} = 2$ keV at $1.5''$ (i.e. 0.06 pc). 
Next, the modeling of emissivity profile in the very long {\it Chandra} 
data \citep{shcherbakov2010}
suggest $n_{\rm e} = 90$ cm$^{-3}$, $T_{\rm e} = 2.5$ keV at this distance. Farther away, the
density and temperature are lower: $n_{\rm e} = 27$ cm$^{-3}$, $T_{\rm e} = 1.3$ keV at about
$1 $ pc \citep{baganof2003}. On the other hand, from the spectral fitting of 
$134.77$ ks {\it Chandra} observation, \citep[][in preparation]{rozanska14}  
estimated the electron temperature of the hot ionized gas to
be $T_{\rm e}= 3.5$ keV at $5''$ (0.2 pc).
On moving to the diffuse gas in the Sgr~A East up to 60'' the  average gas
 temperature is
estimated to be $T_{\rm e} = 3.1$ keV. The surface brightness modelling in the vicinity
of the black hole (up to $3''$, 0.12 pc) requires the Bondi accretion flow with 
outer temperature $T^{\rm out}_{\rm e} = 3.5$ keV and the electron
density $n^{\rm out}_{\rm e} = 18.3$ cm$^{-3}$ for the best fit with observed surface
 brightness.
From this study, the density and temperature estimated at the distance of $1.5''$ 
are $ n_{\rm e} = 29.2$ $cm^{-3}$ and 
$T_{\rm e} = 4.8$ keV respectively.

\begin{figure*}
\hbox{\includegraphics[angle=0,width=9cm]{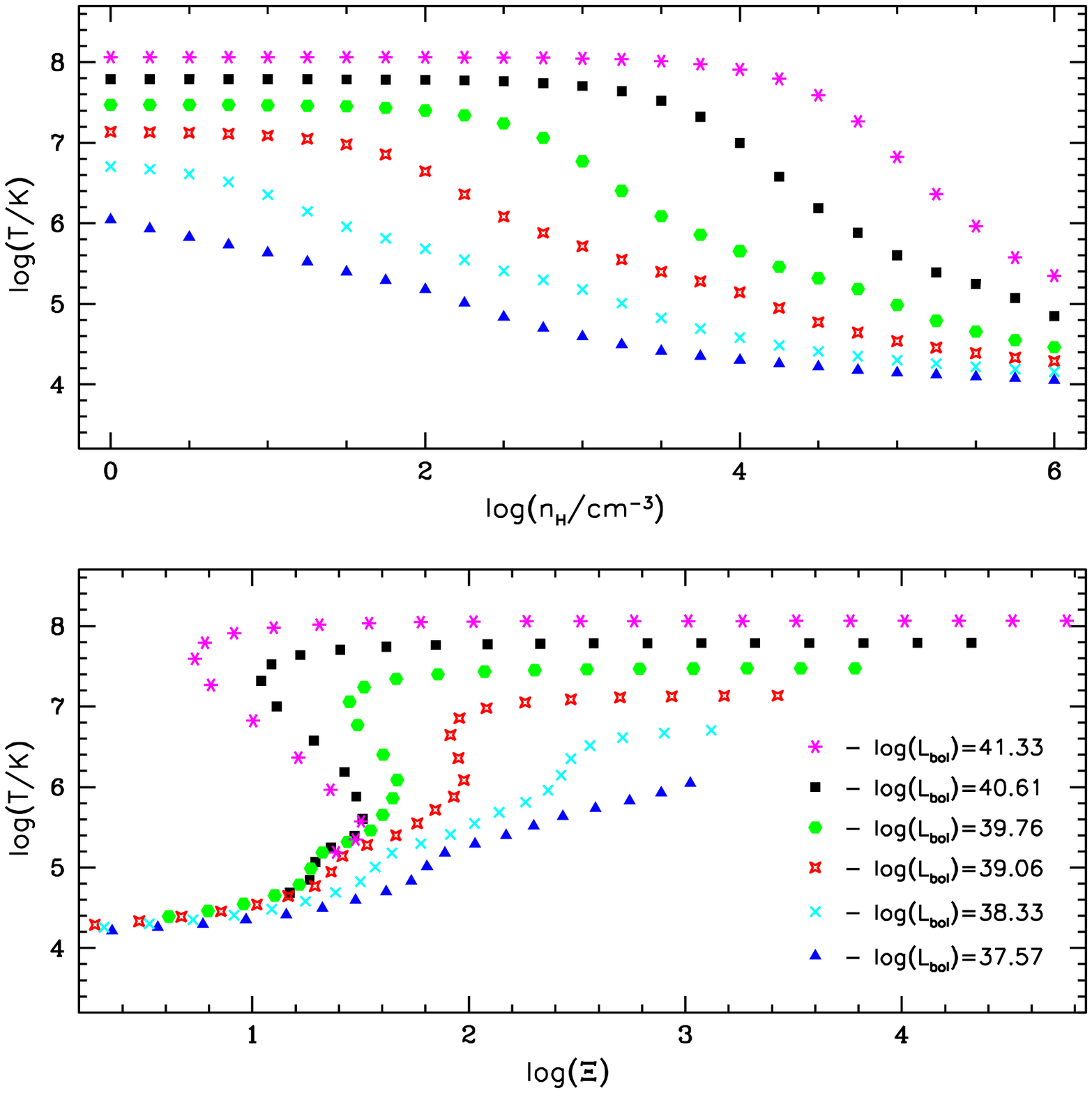}
\includegraphics[angle=0,width=9cm]{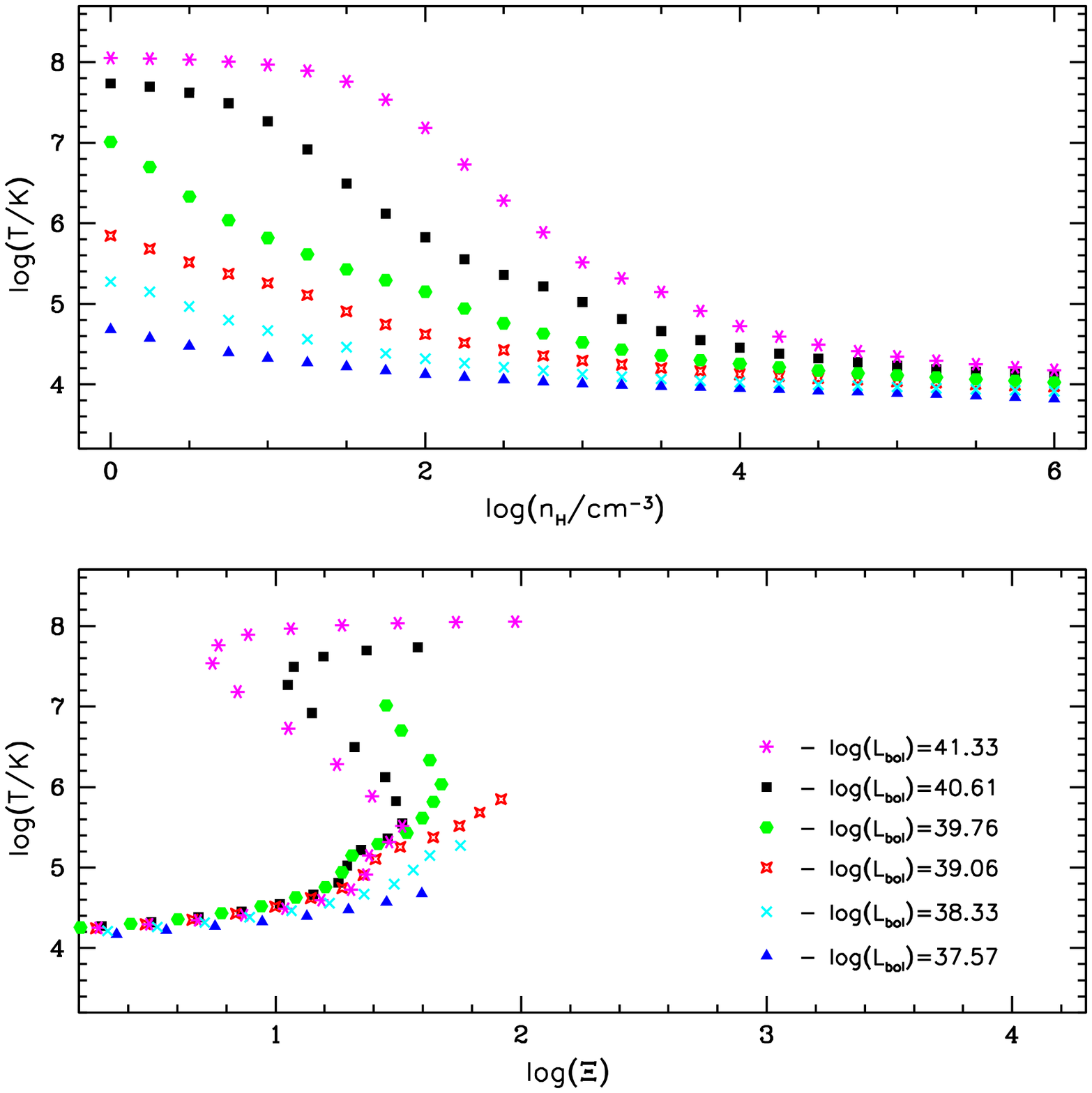}}
\caption{The comparison of thermal instability conditions at two different distances  
from the GC. Left panels show clouds located at $R_{in}=0.008$, while the right panels 
exhibit clouds located at $R_{in}=0.2$.  The instability curves are presented in the 
lower panels, while the temperatures of those clouds versus
their number densities are presented in upper panels for different luminosities.}
\label{fig:xi}
\end{figure*}

The two media (i.e., the hot ionized plasma and the partially ionized gas) appear
to be in mutual contact and, to the first approximation, the pressure equilibrium is
approximately established.  This means that three orders of magnitude higher temperature of
X-ray emitting plasma is compensated by its lower density.  Below we
model a two-phase medium using photoionisation calculations with a proper treatment 
of all cooling and heating mechanisms operating in both types of clouds.

\section{The method}
\label{sec:met}

\begin{figure*}
\hbox{\includegraphics[angle=0,width=9cm]{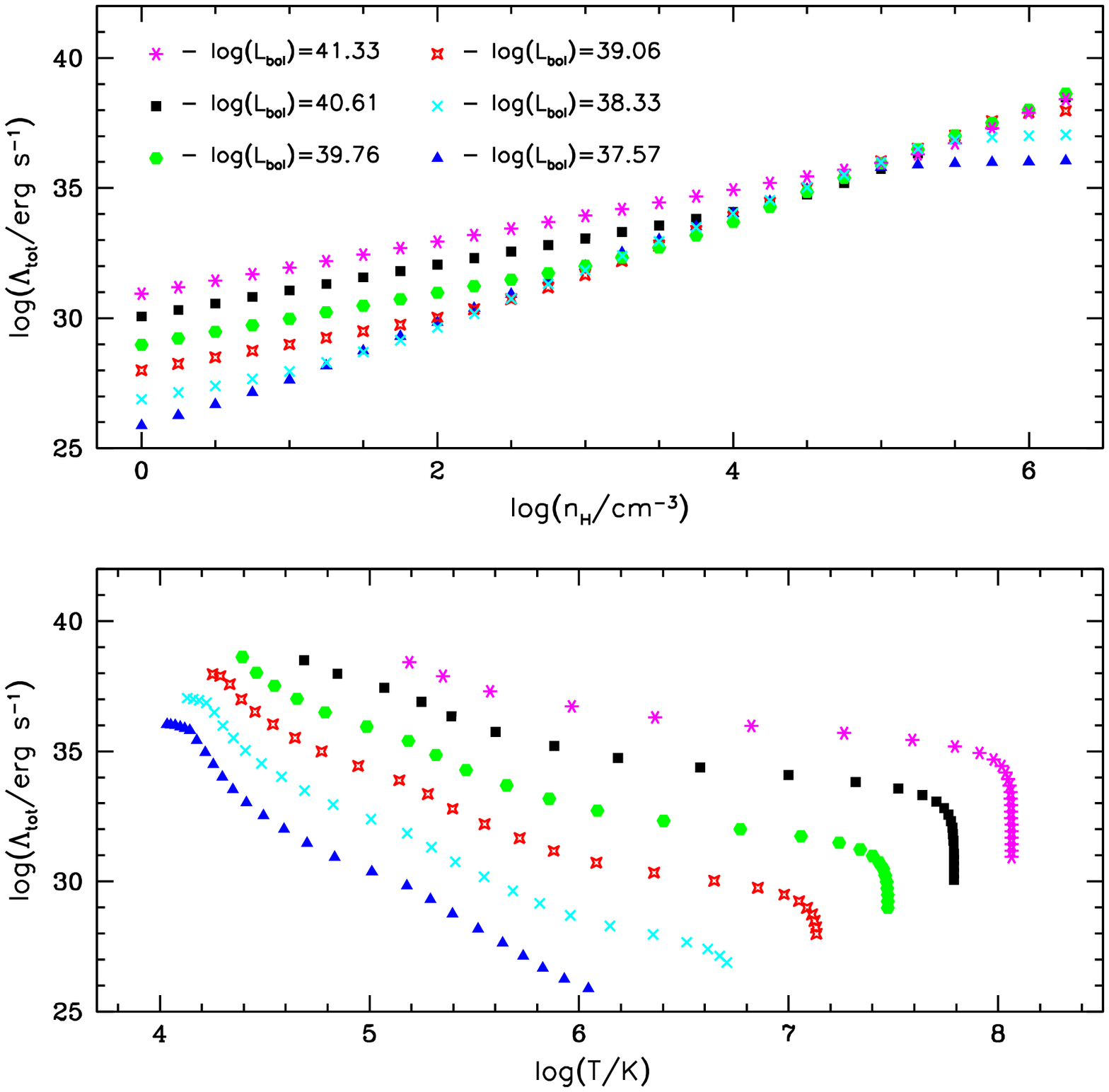} 
\includegraphics[angle=0,width=9cm]{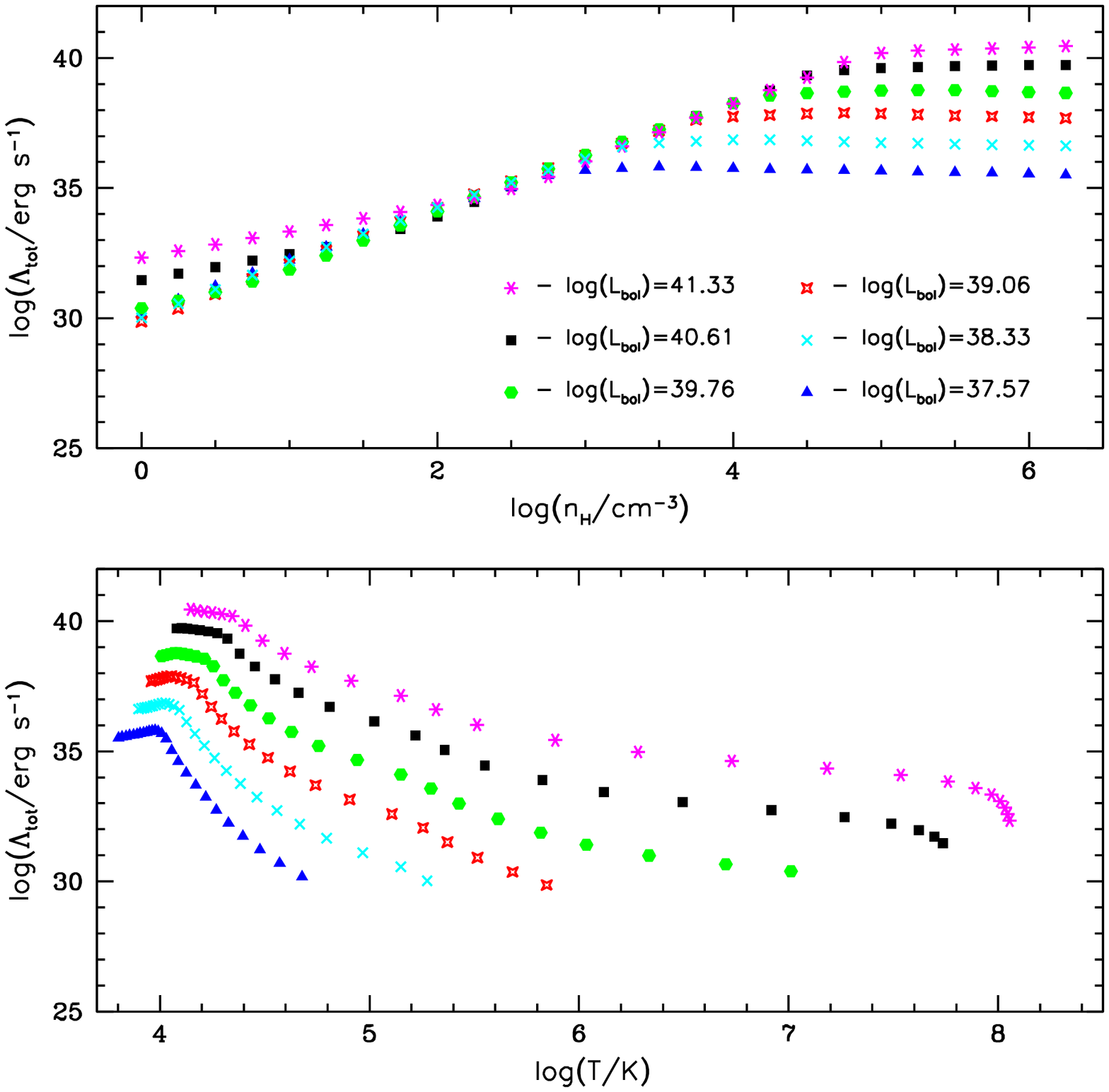} }
\caption{The comparison of total cooling rate of clouds at two different distances,
as in Fig~\ref{fig:xi}. Left panels show clouds located at $R_{in}=0.008$, while right panels
clouds located at $R_{in}=0.2$.  The total cooling rate versus temperature 
for different luminosities is presented in lower panels, while 
the dependence on cloud number density is presented in upper panels for both cases.}
\label{fig:heat}
\end{figure*}

The thermal instability created due to radiative cooling and heating by the
central black hole leads to the formation of a two-phase medium in the
mini-spiral region, with the colder clumps of the mini-spiral structure
breaking off and accreting into Sgr~A*.
This phenomenon can be  modelled using 
photoionisation calculations where the radiation field interacts with 
matter due to bound-bound, bound-free and free-free processes. 
Therefore, we perform Cloudy calculations for initial conditions
taken from observations done recently by working instruments.
Cloudy solves the thermal structure, and  emergent spectrum of 
the illuminated gas under the total ionisation equilibrium. Depending on the 
medium temperature and density, all ionisation and recombination processes 
are calculated for assumed abundances. Therefore, we can determine the total heat
accumulated by the cloud due to irradiation, and follow up heating and cooling 
processes important under the given gas physical conditions. 
 
The thermal instability is seen as a negative slope of the stability curve 
log$(\Xi)$ -- log$(T)$ \citep{krolik1981},  where $\Xi$ is the dynamical ionisation
parameter, defined as the ratio of the radiation pressure to gas pressure:
\begin{equation}
\Xi = {L_{\rm ion} \over {4 \pi c R^2 k n_{\rm H} T}} = {P_{\rm rad} \over P_{\rm gas}} , 
\end{equation}
 where $L_{\rm ion}$ is the total ionizing luminosity, $R$ the distance to the 
source,  $k$ is the Boltzmann constant, $c$  the velocity of light,  
$n_{\rm H} $ hydrogen number density,
and $T$ is the temperature of the ionised medium.

The instability occurs if there is a range of values 
of  the ionisation parameter $\Xi$ for which 
the radiative balance equation has three different solutions of temperature.
This can be illustrated by repeating calculations for 
clouds of different initial number densities $n_{\rm H}$ illuminated by the same 
radiation field. 

We consider clouds of densities spanning from 1 up to $1.78 \times 10^6 $ cm$^{-3}$
(from 0 to 6.25 in log scale).  Each cloud
is located in the inner region of mini-spiral, which we estimate after 
\citet{zhao2010} to 
be $R= 0.2 $ pc (log$(R/{\rm cm})=17.79$) from GC. 
Assuming that the two-phase flow continues toward black hole, we consider also the 
case where clouds are closer at $R_{\rm in}= 0.008 $ pc (log$(R_{\rm in}/{\rm cm})=16.35$) 
toward Sgr~A*. 

For each cloud, we assume open geometry, which means that the cloud is a thin shell 
with  size  $\Delta R$, defined as $\Delta R/ R =< 0.1$ in Cloudy options.
In all calculations, we assume solar default abundances described by 
hazy1.pdf\footnote{http://www.nublado.org/} 
documentation file. It was shown by \citet{hess1997} that the unstable 
branch of stability curve is bigger for higher metal abundances. Especially, 
more iron is responsible for larger extent of the upper part of the stability curve, 
while  more oxygen increases the thermal instability in the lower part of stability curve. 
Nevertheless, for the purpose of this paper we keep solar 
abundances.

For photoionisation calculations, the value of  bolometric luminosity and the 
shape of illuminated radiation are crucial for the shape of stability curve. Since 
Sgr~A* undergoes different luminosity states we have to consider all of them, 
which we describe in the subsection below.

\subsection{Luminosity states of Sgr~A*}
\label{sec:lum}

\begin{figure*}
\hbox{\includegraphics[angle=0,width=9cm]{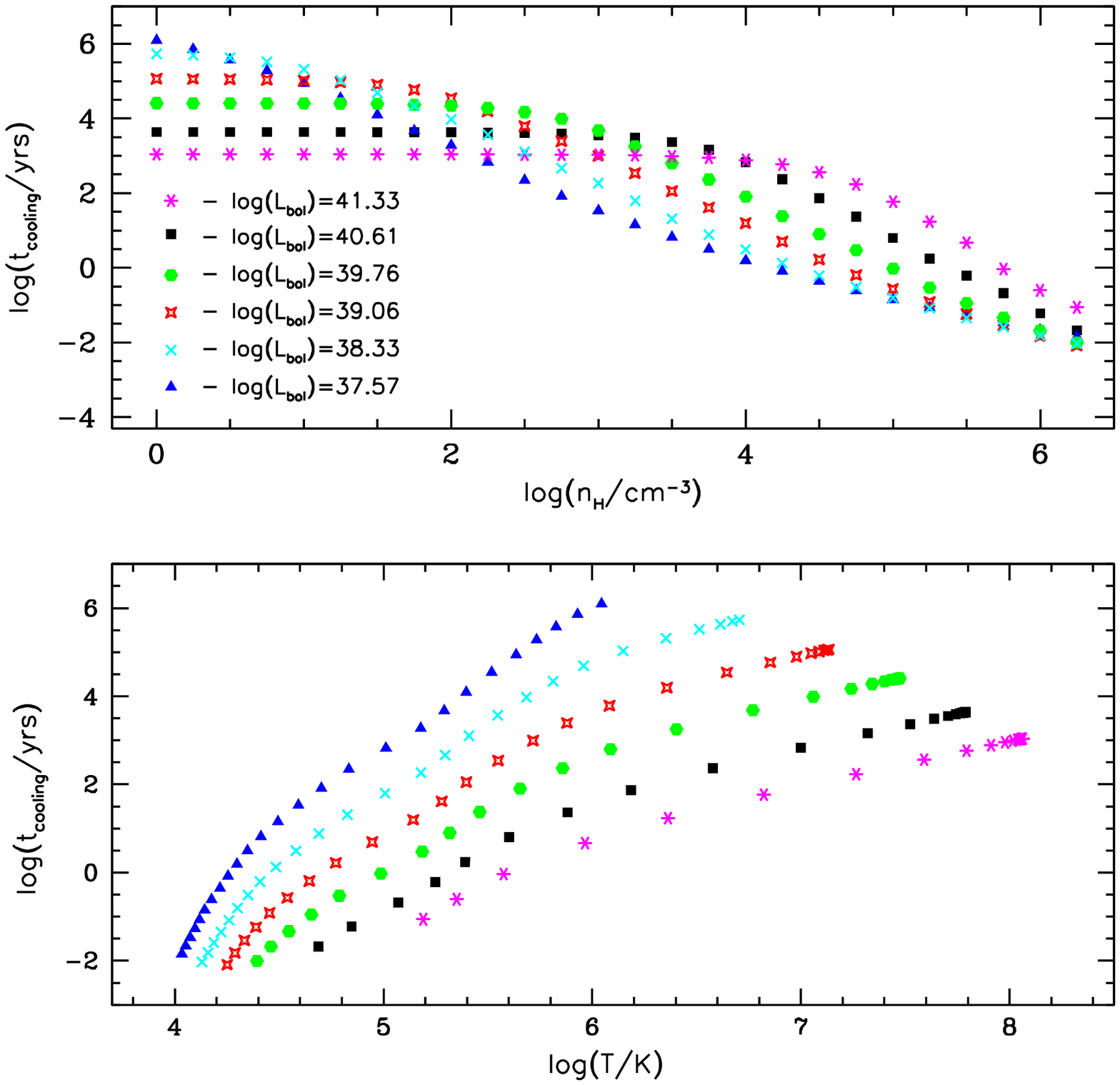} 
\includegraphics[angle=0,width=9cm]{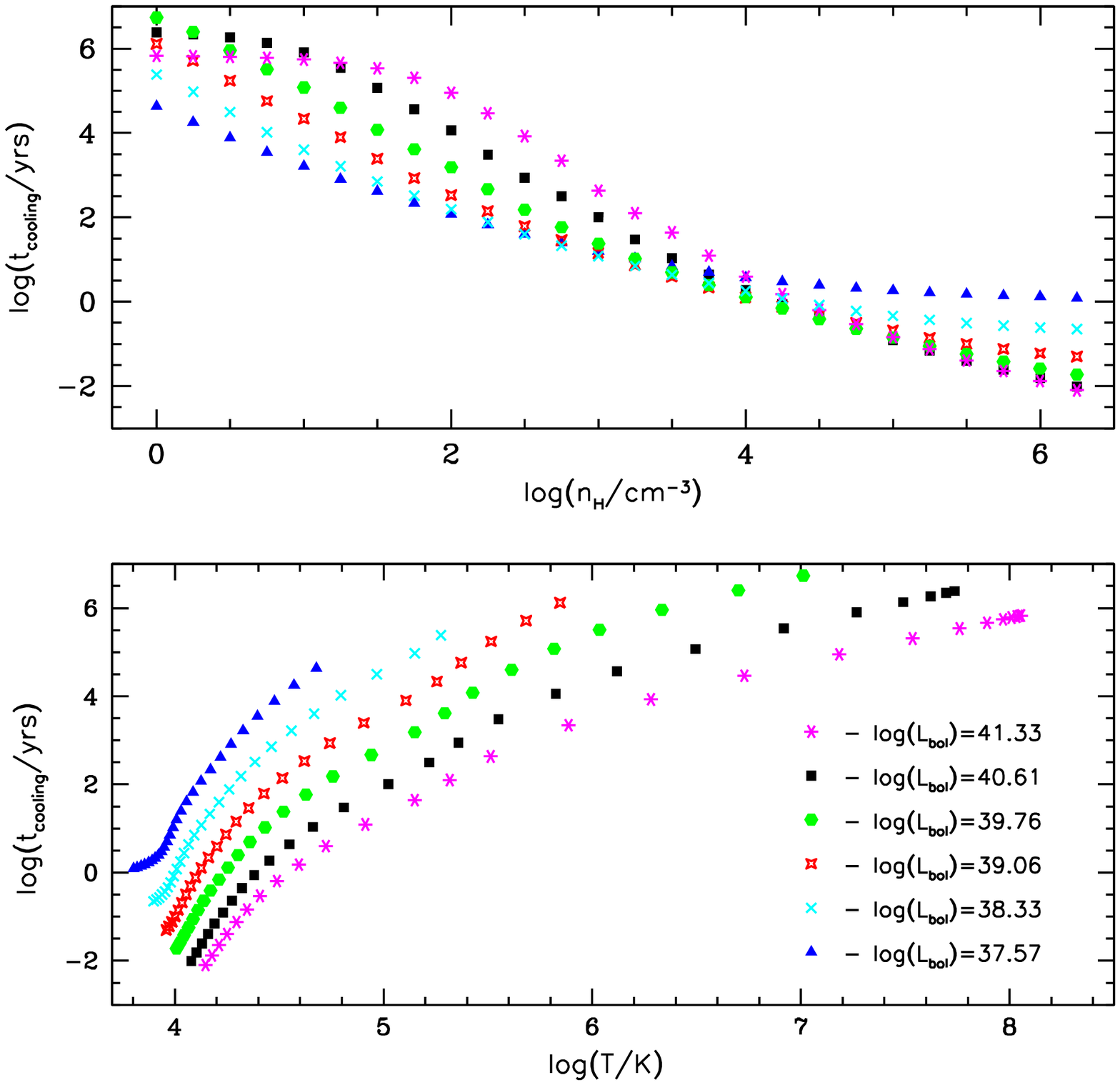}}
\caption{The comparison of cooling time of clouds at two different distances from the GC. 
Left panels show clouds located at $R_{in}=0.008$, while the right panels
clouds located at $R_{in}=0.2$ as in Fig.~ref{fig:xi}.  The cooling time versus temperature 
for different luminosities is presented in the lower panels, while 
the cooling time versus cloud number density is presented in the uppre panels.}
\label{fig:cool}
\end{figure*}

Sgr~A* is highly underluminous, accreting at a rate of about 
10$^{-7}$--10$^{-9}$ times 
the Eddington luminosity. The quiescent X-ray luminosity of Sgr~A* is 
roughly $2 \times 10^{33}$ erg s$^{-1}$, 
while in the flaring or active stage it is $\le 10^{36}$ erg s$^{-1}$. 
The bolometric luminosity is considerably higher due to the strong near-IR 
and mm component. The bolometric luminosity in the past was a few orders
of magnitude higher. Since we do not have the spectral shape of the 
Sgr~A* radiation in the past, we use the theoretical models of GC by 
 by \citet{moscibrodzka2012}  computed for different accretion rates
of radiatively inefficient accretion flow (RIAF).
In those models the black hole mass was fixed at $M_{\rm BH}=4.5 \times 10^6 M_{\odot} $  
\citep{ghez2008}, and distance to the GC at $D=8.4$  kpc \citep{gillessen2009}.
\citet{moscibrodzka2012} computed spectra for accretion rates 
equal to: 1, 2, 4, 8, 16, 32, 64, and 128 $\times 10^{-9}$ $M_{\odot}$  yr$^{-1}$.

In photoionisation calculations we have considered all accretion rates, but 
for the figures presented in this paper we have chosen  six  states of bolometric 
luminosity: from the lowest one log$(L_{\rm bol})= 37.57$, where the instability 
is not present -- to the highest, log$(L_{\rm bol})=41.33$, where the  instability 
is the strongest. Between those two extreme values we present also cases for:
log$(L_{\rm bol})= 38.33$, log$(L_{\rm bol})= 39.06$, log$(L_{\rm bol})=39.76$,
and log$(L_{\rm bol})=40.61$. The shape of each spectrum is taken from the 
typical hot accretion flow onto a supermassive black hole 
\citep[for detailed prescription see][]{moscibrodzka2012}.

The shape of the broad band spectrum for each luminosity state is given by points,
using Cloudy ``interpolate'' command. Based on the output of Cloudy 
calculations we can determine ionization parameter, $\Xi$, 
of each cloud, the total heating {\rm rate} per volume
of the gas $H_{\rm tot}$ in erg s$^{-1}$ cm$^{-3}$, and the total cooling rate
$\Lambda_{\rm tot}$  in erg s$^{-1}$, where the last value allows us 
to determine the cooling time  according to the formula:
\begin{equation}
t_{\rm cooling} = {P_{\rm gas} \over \Lambda_{\rm tot}} = {k n_{\rm H} T \over \mu
 \Lambda_{\rm tot}},
\label{eq:cool}
\end{equation} 
where $\mu$ mean molecular weight = 0.5, as usual. Cloud temperature, $T$, 
is always self-consistently computed from photoionisation calculations.

Thermally unstable clouds would evaporate or condensate due to thermal instability 
and thermal conduction.  If the heating/cooling
 curve shape implies the instability, the condensations will grow out of the 
initially uniform medium. Thermal conduction limits the smallest size of the clouds.
It does not influence the size of thermal instability as it was shown 
by \citet{begelman1990,rozanska99}.
Taking into account classical thermal conductivity based on the assumption that 
the mean free path is short with respect to the temperature scale height, for plasma
of cosmic abundance the conductivity is equal to $\kappa = 5.6 \times 10^{-7} T^{5/2}$
erg~cm$^{-1}$~s$^{-1}$~K$^{-1}$ \citep{draine84}. The size of such clouds 
can be estimated by Field length \citep{field1965} which is a ratio of 
mean conductive flux to the total heating  rate per volume:
\begin{equation}
\lambda_{\rm F} = \left( {{5.6 \times 10^{-7} T^{7/2}} \over {H_{\rm tot}}} \right)^{1/2} ~\ .
\label{eq:field}
\end{equation}

The formation timescale of a cloud of this size is well approximated by 
Eq.~\ref{eq:cool}
if the sound-crossing timescale across the cloud is much shorter than 
$t_{cooling}$  \citep{field1965,burkert2000}.
The cloud size equal to Field length implies that the role of radiative
 cooling and conduction is comparable. In this case, if the external 
heating/cooling switches off, the cold cloud surrounded by the hot
 plasma will disappear due to the conduction in the same timescale.

On the other hand, we can simply estimate the total mass stored in the cloud:
\begin{equation}
M_{\rm C} = m_{\rm H}  n_{\rm H}  V = m_{\rm H}  n_{\rm H} 
 { \Lambda_{\rm tot} \over H_{\rm tot} }    ~\ ,
\label{eq:mass}
\end{equation}
where $m_{\rm H}$ is the mass of hydrogen atom, and $V$ is the physical volume 
of the cloud in cm$^3$. 

\section{Thermal instabilities for different states of luminosity}
\label{sec:res}

As a result of Cloudy computations we present the stability curve,
 for different luminosity states discussed above.
In Fig.~\ref{fig:xi}, lower panels, we present the comparison of
stability curves for clouds located 
at  two different distances, 0.008 and 0.2 pc from Sgr~A*. 
At the closer distance, 
the instability is not present if the luminosity 
of the central source is too low, log$(L_{\rm bol}) < 39.06 $. However, it operates
for four examples of high luminosities considered in our calculations. 
Further away from the GC, the instability is present only for 
the two highest luminosities, Fig.~\ref{fig:xi} (lower right panel).

From the upper panels of Fig.~\ref{fig:xi}, where we present the 
dependence of cloud temperature versus their number density,  we see that unstable 
clouds are denser when they are located closer to the Sgr~A*. 

The comparison of total cooling rate of clouds at two different distances from the GC 
is shown in Fig.~\ref{fig:heat}.
The total cooling rate of each cloud is higher for higher cloud number density
(upper panels) and for lower cloud temperature (lower panels). 
Additionally,  the total cooling rate is much higher for clouds
located farther away from GC (upper right panel of Fig.~\ref{fig:heat}).
The dependence of total cooling rate on number density 
exhibits the pivoting point, which occurs at different number density, 
depending on the cloud location. However, it does not seem to have any special 
physical consequence that could be checked observationally.
We address this issue for further theoretical consideration in the future.

Finally, we show in Fig.~\ref{fig:cool}, how total cooling time computed from 
Eq~\ref{eq:cool}, depends on density and temperature. Typically, clouds located
at 0.008 pc from Sgr~A*, and with  $n_{\rm e} < 3 \times10^4$ cm$^{-3}$ can
survive more than 100 years (upper left panel of  Fig.~\ref{fig:cool}).  
The cooling time is longer for higher luminosity state, 
but for high temperature clouds it saturates at the lower value. 
For the log$(L_{\rm bol})=37.57$, the coolest cloud has the highest possible lifetime
$t_{\rm cooling}=10^6$ yrs.  For more distant clouds, their lifetime is shorter,
for the lower luminosity states, but longer for the higher luminosity states
(right panels of Fig.~\ref{fig:cool}).
Still clumps at $n_{\rm e} < 10^3$ cm$^{-3}$ can survive more then 100 years.

Comparing this approximate cloud cooling time to the timescale of free fall of 
the cloud onto the supermassive black hole in the GC, we can determine the inner 
distance $R_{\rm min}$ from the Sgr~A*, at which $t_{\rm ff}=t_{\rm cooling}$.
\begin{equation}
R_{\rm min} = \left({2 \over \pi} \, * \, t_{\rm cooling} * [\, 2 \, G \, 
(M_{\rm BH} + M_{\rm cl} )]^{1/2} \right)^{2/3}
\end{equation} 
where $G$ is a gravitational constant, and mass of the cloud, $M_{\rm cl}$, 
is negligible in comparison with $M_{\rm BH}$.  
If we are interested in clouds living minimum 100 years,  they should be 
visible at distances equal or higher than $R_{\rm min}=0.054$ pc. It is in good  
agreement with observations of mini-spiral which is visible down to 
$\sim 0.1-0.2$ pc towards the central black hole.
Clouds located closer to the black hole then the calculated 
value of $R_{\rm min}$, fall onto black hole before 
the end of their cooling time. 

The extent of the instability depends on the extension of two-phase medium 
visible on the stability curve. As seen in Fig.~\ref{fig:xi}, lower left panel, 
for the highest 
four luminosity states the instability range equals: 0.77, 0.46, 0.26 and 0.07 
in terms of log$(\Xi)$. 

At the intermediate distance, 0.08 pc, the stability curves for the same 
luminosity states are showing that in case of three highest luminosities 
two-phase medium can arise (Fig.~\ref{fig:dust}, upper panel).
Following stability curves, computed for all three distances, it is clear 
that for the same gas physical parameters, the strongest instability occurs 
closest to the radiation source.

Interestingly, the size of the instability,  i.e. the range of ionisation parameter
where thermal instability operates, 
is the same at all three distances if we
compare the same luminosity states, even if densities are different. 
Therefore, we conclude that particular thermal instability can operate far
away from GC, and create two-phase medium, if the number density  of matter  
allows it. We check this prediction in Sec.~\ref{sec:two}.

\subsection{The size of clouds}

\begin{figure}
\epsfxsize=5.0cm \epsfbox[40 160 350 690]{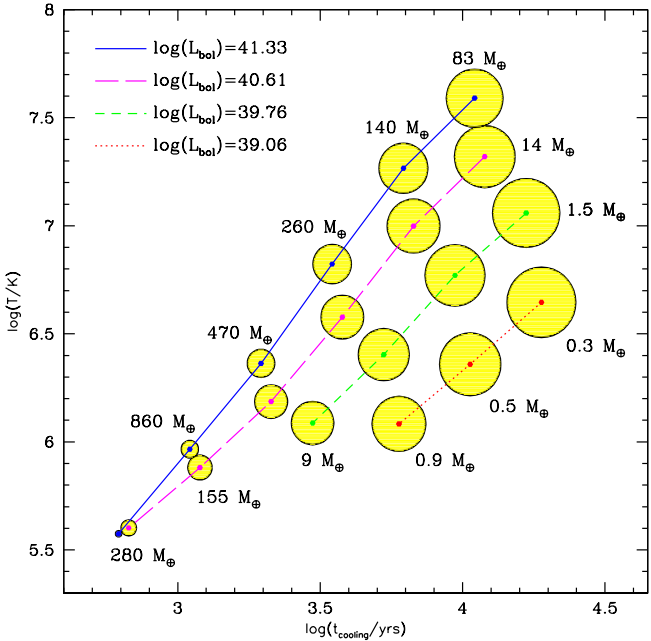} 
\caption{The size of clouds along the instability branch positioned in the 
log-log plot of cooling 
time $t_{\rm cooling}$ vs.\ temperature $T$. Four different curves 
are plotted for different values of bolometric luminosity, i.e., 
the four lines connect clouds that are produced by the same radiation field, 
as specified in the top-left corner of the plot.
 Along each curve, the radius of yellow 
circles represents the Field length according to Eq.~\ref{eq:field}. 
Masses of clouds determined by Eq.~\ref{eq:mass} are given in the units of Earth mass. 
}
\label{fig:clo}
\end{figure}

In Fig.~\ref{fig:clo} we present the overview of  clouds being on unstable branch for 
those luminosity states where instability occurs. Depending on luminosity, clouds 
have different sizes and masses. We connected by lines those clumps which are 
produced by the radiation field of the same intensity.   

The size of the circles is given by the Field length of each cloud (Eq.~\ref{eq:field}),
and since there is a difference by many orders of magnitude we presented it
 in logarithmic scale. 
The largest clouds have a radius of the order of $10^{17}$ cm, and they appear either for 
low luminosity states or for high temperature i.e. low density clouds. Those clouds have 
the smallest mass reaching even $ 0.3 M_{\oplus}$ .
The heaviest compact clouds are created for low temperatures and high densities 
of the order of  $10^5$ cm$^{-3}$. They have radii approximately $10^{12}$ cm,
and can accumulate huge mass of the order of few hundred Earth masses. 
 
Unstable clouds with mass of about 10 $M_{\oplus}$ can exist even more than ten
thousand years. Those are very good candidates for a G2-type cloud in agreement with
the recent mass estimation  of 4 -- 10 $M_{\oplus}$ given by \citet{shcherbakov2014}. 

\subsection{The effect of the dust}

Strong extinction towards Sgr~A* has been interpreted as a result of the
line-of-sight absorption by the intervening material with a significant
contribution from interstellar dust grains \citep{krishna2005}. Dust in
the GC can be formed mainly in atmospheres of cool stars
through the process of mass loss and via supernovae explosions that have
occurred in the region and helped to disperse the grains into the ISM.
Alternatively, the dust can also originate at larger distances from the
centre, where the conditions for nucleation are favourable and the
material can be transported towards the centre in filaments and clouds.
However, these must interact with the wind from hot He stars and the
likely wind from Sgr~A*. Therefore, the role of dust should be explored
in the context of the proposed thermal instability and the structure
formation near the supermassive black hole.

Order-of-magnitude estimates indicate the typical interstellar
dust-to-gas ratio about $\sim1$\% is maintained. Complexities of the
thermal structure and stratification in the GC region have
been revealed by infrared absorption features \citep[e.g.][ based on ISO satellite observations, and based on ESO's VLT $M$-band spectra, respectively]{gibb2004,moultaka2009}.
These have been identified as shocks and filaments
that develop in the region. Bow shocks presumably originate from the
interaction of stars embedded in the ISM of the GC
\citep{clenet2004,muzic2010}, which points to the importance of stars
as an additional source of energy. Therefore, conditions for dust
formation are particularly varied through the central region.

\begin{figure}
\epsfxsize=5.0cm \epsfbox[40 160 350 690]{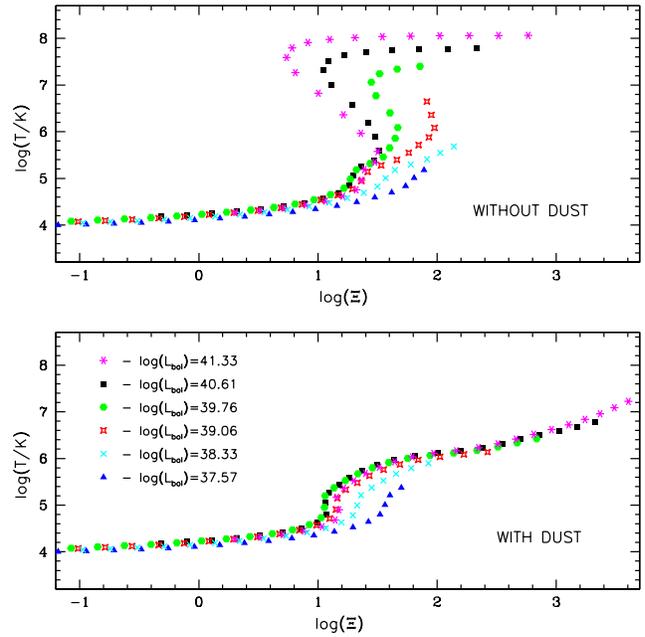} 
 \caption{The comparison of stability curves computed without dust and with dust
 for clouds of the same physical parameters and for the same luminosity
states as in previous figures. All clouds presented at upper and lower panels are located at moderate distance 0.08 pc from GC. }
\label{fig:dust}
\end{figure}

The thermal instability is not the only mechanism that can enhance
mixing and, consequently, drive accretion in the complex environment.
Moreover, the abundance of heavy elements generally increases towards
the GC, so a higher dust-to-gas ratio can be expected. This
raises a question whether the presence of dust could suppress the
instability close to the black hole by acting as an efficient heat sink.
On the other hand, dusty plasma is known to become unstable by a variety
of instability modes that are driven under external action, depending on
the actual chemical composition and size of dust grains 
\citep[see][and further references cited therein]{mendis1994,cox2005,vladimirov2005,zajacek2014}.
 In addition, irradiated by UV/X-ray photons and embedded in
the plasma dust grains must acquire a significant electric charge
\citep{ishihara2007}. The presence of charged dust in the plasma modifies the
usual instabilities and the electrostatic repulsion can even contribute
to erosion and rapid destruction of dust.

In our computations discussed in the previous section we neglected the existence of dust. 
Observations show clearly that, at present, the mini-spiral also contains  dust
\citep{muzic2007}.
 Therefore, we performed the computations of the thermal instability 
log$(\Xi)$ -- log($T$) curve  with the dust 
grain option included in Cloudy, assuming the standard dust 
composition characteristic for the interstellar medium. 
The effect is illustrated in Fig.~\ref{fig:dust}, where we compare the 
stability curves for the same physical parameters with dust (lower panel), 
and without dust (upper panel) at a moderate
distance of 0.08 pc, i.e. $2''$ or log$(R)=17.39$ cm, from GC.

\begin{figure}
\epsfxsize=5.0cm \epsfbox[40 160 350 690]{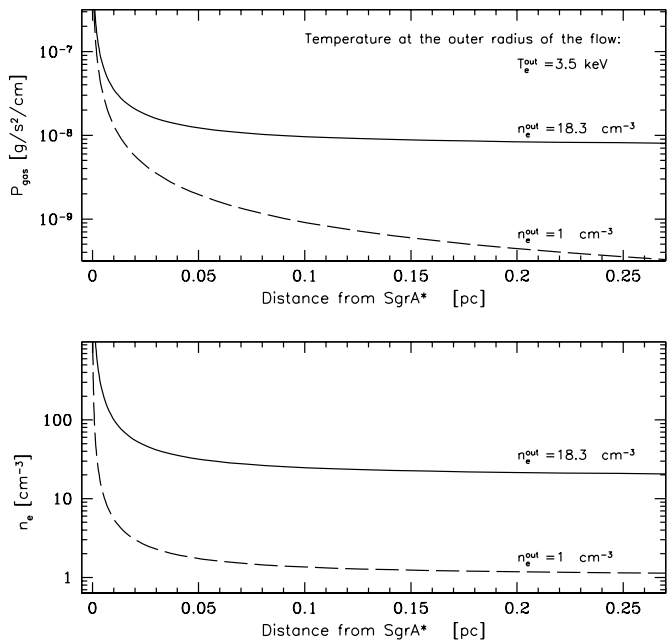} 
 \caption{The gas pressure profile (upper panel), and density profile (lower panel) 
for hot phase, in case of Bondi accretion flow with $T^{out}_e=3.5$ keV as it was found 
in X-ray data. Two values of outer number density are considered. 
From the best fitting model $n^{out}_e=18.3$ cm$^{-3}$ -- solid line,
and for the large extent of instability $n^{out}_e=1$ cm$^{-3}$ -- dashed line. }
\label{fig:gas}
\end{figure}

The low temperature branch, with the temperatures well below $T \sim 10^5$ K is unaffected 
by the presence of the dust. High temperature part of the diagram is strongly modified: the
 gas temperature is much lower since the dust intercepts significant fraction of the 
incident radiation. 
The dust temperature, even for the very high luminosity, is still
 moderate, $T_{\rm dust} \sim 300$ K so the high level of Sgr~A* activity does not destroy 
the dust radiatively. However, it is well known that hot surrounding plasma easily 
destroys the dust particles collisionally \citep[e.g.][]{draine1979}. In particular,
 recent computations of the dust particle erosion in a hot gas by 
\citet{bocchio2012,bocchio2013} indicate the destruction timescale of 10 years 
for a hot plasma number density of 10 cm$^{-3}$.
 This means, that the dust survives only within the warm medium clouds, where it does not
 affect the temperature much, and vanishes immediately from the hot phase. 
Thus, the stability curve for the dustless plasma applies for timescales larger than a year.
 On the other hand, on shorter timescales the effect of dust is visible
\citep{krabbe95} 
especially if its efficient formation in stellar winds replenishes the dust 
content at a sufficiently fast rate to ensure the steady state.

\section{Two-phase medium in Bondi flow around Sgr~A*}
\label{sec:two}

\begin{figure*}
\hbox{\epsfxsize=9.0cm \epsfbox{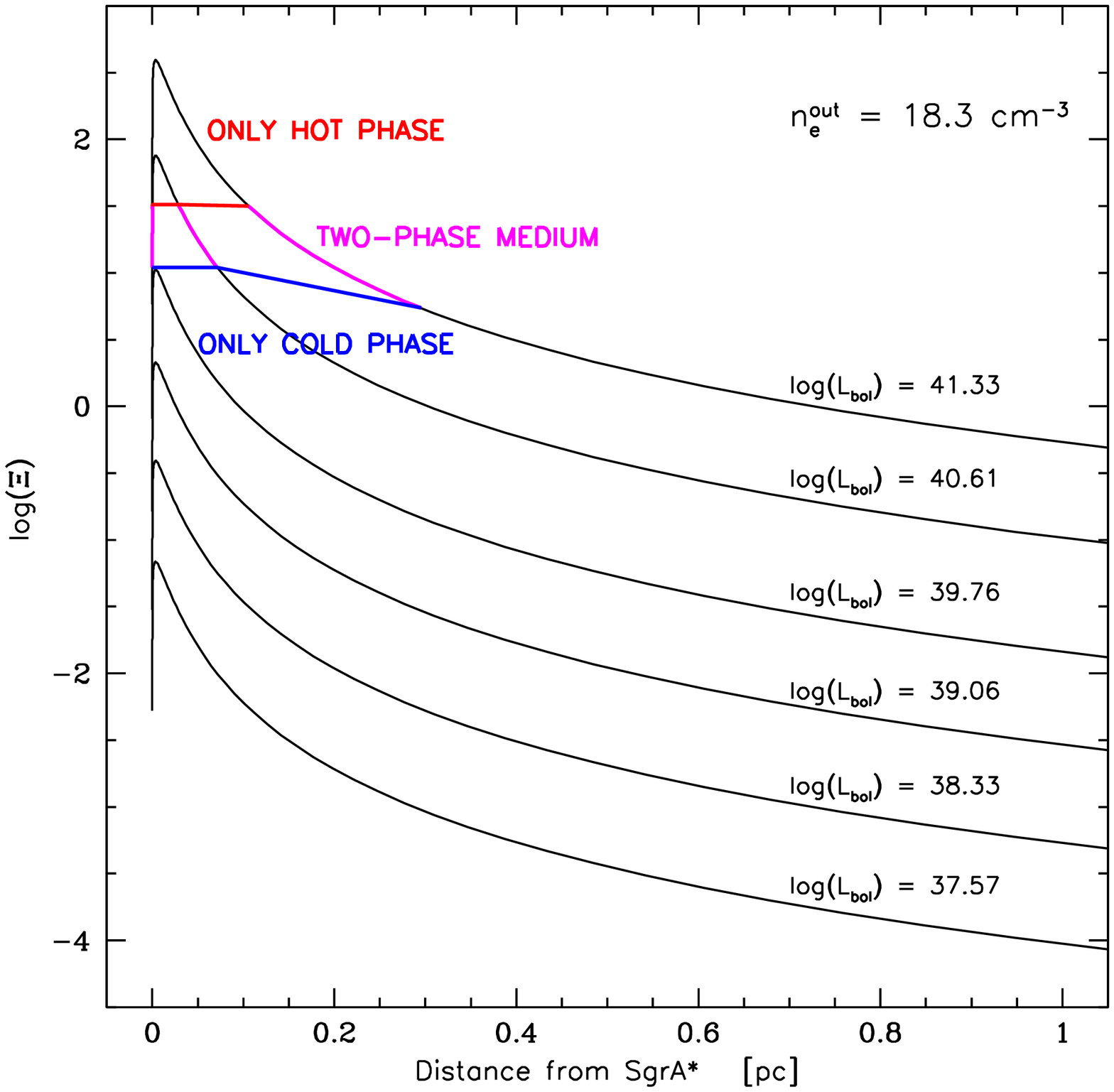} 
\epsfxsize=9.0cm \epsfbox{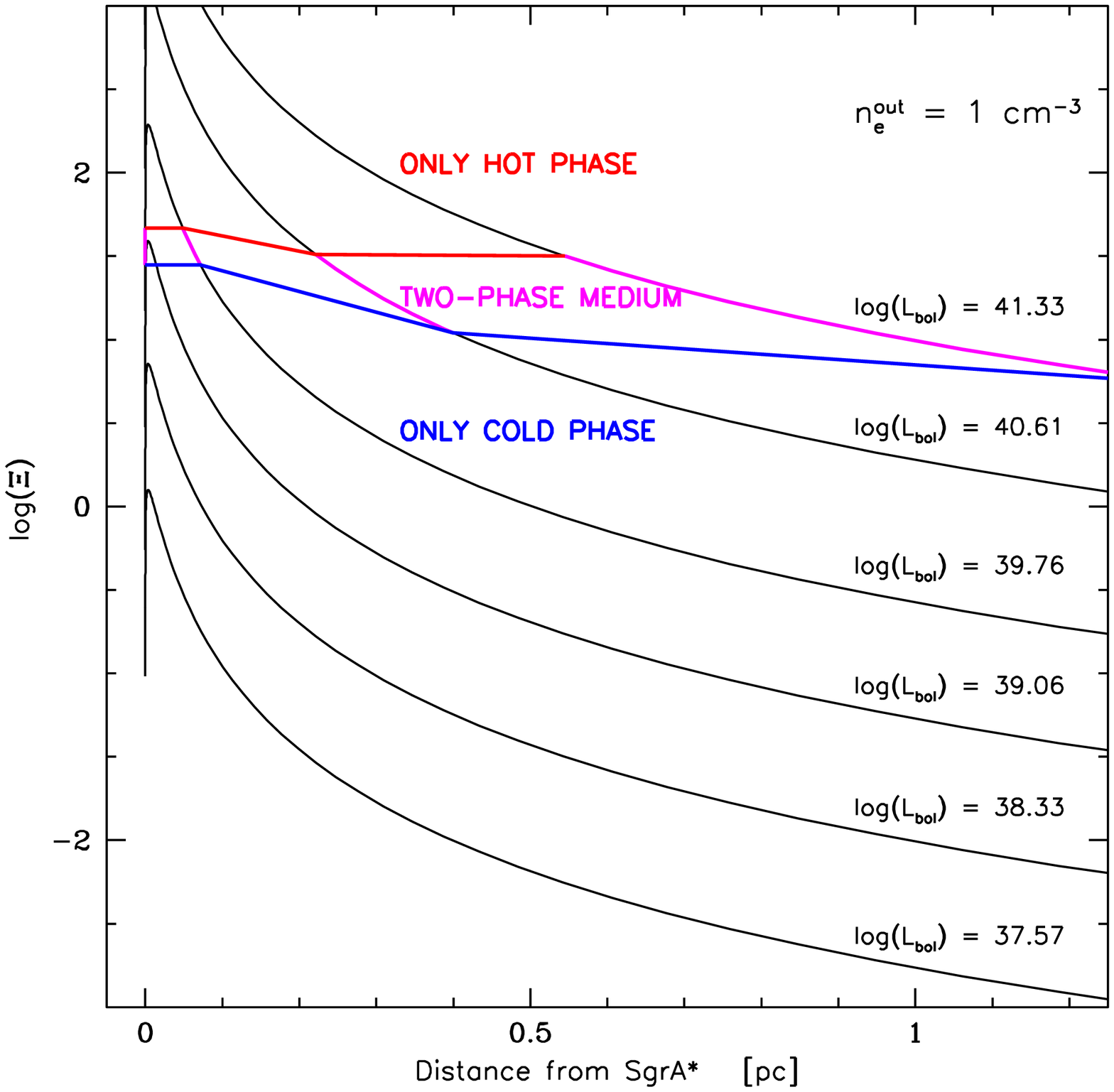}} 
\caption{Instability strips for the different luminosity states in case of Bondi 
accretion flow onto Sgr~A* for outer temperature $T^{out}_e=3.5 $ keV. 
Left panel shows cases for outer number density $n^{out}_e=18.3$ cm$^{-3}$ and 
right panel for $n^{out}_e=1$ cm$^{-3}$. 
The region in between colour contours is the  two phase-medium, and its extent is shown
by magenta lines for each luminosity state. The thermal instability can operate
in range of high luminosity for Sgr~A*, which occurred in its past history.}
\label{fig:ins}
\end{figure*}

From the fitting of X-ray data surface brightness profile, it was found that 
Bondi accretion operates up to 0.12 pc from Sgr~A* \citep[][in preparation]{rozanska14}. 
Further away hot plasma is interacting with stars and the more advanced dynamical model, 
where Bondi flow is mixed with stars, fits better
as shown by \citet{shcherbakov2010}, where fit to {\it Chandra} data was
done up to 0.2 pc.

For this paper, we adopt that the hot plasma in the mini-spiral region
is well represented by Bondi accretion flow.
We assumed here, that the inner hot accretion flow,
for which luminosity states were computed (Sec.~\ref{sec:lum}) changes to the 
Bondi flow for larger distances with zero angular momentum. 
This is in agreement with the work done by \citet{cuadra2008}, where 
the circularized radius was estimated on $R_{\rm circ} \sim 0.002$ pc.  
For the Bondi accretion flow starting roughly at the capture radius, 
we were able to compute density  and gas pressure 
profile around Sgr~A*, with initial parameters $T^{\rm out}_e=3.5$ keV,
and  $n^{\rm out}_e=18.3$ cm$^{-3}$, defined on the outer radius of the flow (far above 2 pc)
taken from observations \citep[][in preparation]{rozanska14}.

In Fig.~\ref{fig:gas}, we present the gas pressure and the number density profiles 
versus distance from the black hole for the Bondi accretion. 
We consider one outer temperature 
and two outer number densities: $n^{\rm out}_{\rm e}=18.3$ cm$^{-3}$, and 
$n^{\rm out}_{\rm e}=1$ cm$^{-3}$. 

Finally, Fig.~\ref{fig:ins} represents instability 
strips computed for both models respectively. For each luminosity state
 at each distance we calculated the value of $\Xi$ parameter taking 
into account gas pressure and density 
number profiles for both cases of Bondi flows.
The instability strips are marked as magenta colour for each luminosity state. 

The occurrence of instability during the high-luminosity
history of Sgr~A*, and its disappearance at the present low-luminosity stage,
represent the main argument of this paper. 
For all gas above the red contour, only hot phase can exist at a given luminosity. 
For the matter below the blue contour, only cold clouds can exist. 
There are several distance ranges where two-phase medium can co-exist. 
Nevertheless, for Bondi flow with  $n^{\rm out}_{\rm e}=18.3$ cm$^{-3}$ 
(left panel of Fig.~\ref{fig:ins}) it appears only for 
two highest luminosities and  quite close to the Sgr~A* within 0.4 pc. 
In case of Bondi flow with $n^{\rm out}_{\rm e}=1$ cm$^{-3}$ (right panel of Fig.~\ref{fig:ins})
two-phase medium can be sustained up to 1.4 pc.

\section{Discussion}
\label{sec:dis}

Our estimates presented in this paper are clearly oversimplified, 
since we did not include the additional
heating of the medium by stars, which 
explains observed X-ray brightness profile above 
0.12 pc \citep{shcherbakov2010}. Such an additional heating may slightly 
shift the instability region but in general it enhances the instability.

The multi-phase character of the medium affects to some extent the accretion 
pattern, as showed by \citet{barai11,barai12}, and \citet{moscibrodzka13}.
The main effect can be easily seen from consideration of the simplest 
spherically symmetric 
Bondi accretion flow. The Bondi radius for the hot plasma is of order of 
\begin{equation}	
R_{\rm Bondi}^{\rm hot} = 0.074 q_{\rm s} {M_{\rm BH} \over 4.5 \times 10^6 M_{\odot}} 
\mbox{[pc]},
\end{equation}
where the sound speed for the hot material has been set to 500 km s$^{-1}$ and $q_{\rm s}$ 
coefficient depends on the adopted polytropic index $\gamma$. If $q_{\rm s} = 1$, the inner 
 edge of the mini-spiral is close to that radius. However, for 
$\gamma = 1.6$, $q_{\rm s} = 0.05$
 and the hot material at the inner radius of the mini-spiral is not yet flowing in. 

For the cold material, the square of the sound speed is by a factor of $10^3$ lower 
so the infall is 
possible even from larger distances than the inner edge of the mini-spiral, as the 
formal Bondi radius for would be three orders of magnitude larger
\begin{equation}
R_{\rm Bondi}^{\rm hot} = 3.7 {M_{\rm BH} \over 4.5 \times 10^6 M_{\odot}} \mbox{[pc]}.
\end{equation}
Here we adopted $\gamma = 1.6$. Therefore, the cold material falls immediately 
into the black hole as soon as it looses its angular momentum.
This is exactly the pattern seen in the numerical computations: cold clumps may
accrete when the hot medium which surrounds them, is outflowing due to the strong heating.

The dynamics of the mini-spiral is even more complicated since its material
possesses a significant amount  of angular momentum  \citep{zhao2009}. 
However, clouds can loose angular momentum in the collisions. 
The mutual collisions lead to 
the gradual dissipation of angular momentum, so that the individual clouds are
 expected to fall more rapidly to the center. Also the radiation drag by the 
central source and the luminous inner accretion disc act on the cloud to change
 the ratio of azimuthal versus radial velocity components. For clouds of sufficiently
 large optical thickness, the shape and stability of their orbits is strongly 
affected by the radiation pressure especially during the luminous phase of the 
centre \citep[e.g.][]{fukue99,ghisellini04}. The radial 
component of the motion then starts to grow relative to purely angular 
component, as was studied, including the relativistic effects acting on the 
motion \citep{abramowicz90,vokrouh91}. The properties of 
particle motion can be employed to effectively  represent 
the cloud mean motion, however, different aspects 
(such as the role of the cloud internal temperature) 
have been also discussed \citep{keane2001,horak06}. 
It has been shown that in case of more general (non-radial) orbits, although
 the shape of bound trajectories deviates from ellipses, the motion remain 
closed \citep{plewa13}. Therefore, we assume that non-gravitational
 effects contribute to higher-order corrections and remain beyond the scope of 
the present paper.
Furthermore, \citet{cuadra2005} have studied the expected properties of the 
process accretion of cool stellar winds on to Sgr A* supermassive black hole. 
These authors notice that cool streams of gas frequently enter this region on
 low angular momentum orbits. Then the streams are disrupted and heated up to 
the ambient hot gas temperature.

Cloudy photoionisation computations allow to search for the influence of 
dust and cosmic rays on the thermal instability. The dust was extensively 
discussed in the previous section. We have made test on how typical cosmic 
ray background can change the shape of stability curve, and we did not observe
even small effect.

It was shown in the original paper \citep{field1965} that the presence of the 
magnetic field penetrating the electrically conducting gaseous medium leads to several 
different effects that can modify the conditions for the thermal (non-gravitational) instability. 
Firstly, the internal pressure gets increased by the additional magnetic influence, 
so the pressure term must be replaced by the sum of gas plus magnetic terms 
\citep{langer78}. Next, the presence of the magnetic field leads to 
an anisotropic heat flow since the motion of electrons is greatly reduced in the direction 
across the field lines. However, in the plane-parallel approximation the growth timescale 
of the condensation instability is not diminished strongly, unless 
the magnetic field is strictly perpendicular to the temperature 
gradient \citep{field1965,vanhoven84}.
 This may not be true in a complete (three-dimensional) picture since the 
magnetisation will 
 likely lead to the creation of complex filamentary structures. Hence, advanced numerical 
 simulations must include this anisotropic term \citep{sharma2007,parrish09}. 
Furthermore, suppression of thermal conductivity in the direction 
normal to field lines depends on the state of the medium, i.e., the case of fully 
vs.\ partially ionized gas.
Furthermore, \citet{burkert2000} found that tangled magnetic fields can reduce the
 conductive heat flux enough for low-amplitude fluctuations to grow and become nonlinear 
if their length scales are of order $\sim0.01$~pc. These authors argue 
that if the amplitude of the initial perturbations is a decreasing function of the
 wavelength, the size of the emerging clumps will also decrease with increasing strength 
of the magnetic field.

Therefore, the clumpiness of the mini-spiral medium caused by the thermal instability 
creates a possibility of episodes with enhanced accretion of cold clumps towards Sgr~A*
and may well explain its enhanced activity in the past.

\section{Conclusions}
\label{sec:concl}

The central region of the Milky Way consists of the supermassive black hole, a 
hot X-ray emitting plasma filling roughly uniformly the space, and the 
filaments of the colder gas and dust. Some of these filaments, in the form of  
the mini-spiral, extend down to $\sim 0.1-0.2$ pc towards the central black hole.
Here we analyzed a possibility that the two-phase medium 
in the mini-spiral region close to Sgr~A* might have formed due to the thermal 
instability which can develop in the irradiated medium.

We showed that the current level of the X-ray emission from the vicinity of Sgr~A*
is not high enough to drive the thermal instability. However, there are strong 
arguments that Sgr~A* was orders of magnitude brighter in the past, about a 
hundred years ago. In that case, for central luminosities higher than 
$\sim 10^{39}$ erg s$^{-1}$, in the medium at distances 0.008 - 0.2 pc, 
 the instability operates and the 
heated medium spontaneously forms colder clumps embedded in the hot low density
plasma. Such two-phase medium exists for hundreds of years after the turnoff
of the nuclear emission since the cooling timescale of the hot gas is very long.
We have shown that clouds located at distance higher than 0.054 pc from Sgr~A*, 
can survive hundreds years, since this cooling/evaporation time is shorter than 
free fall time at this distance. Clouds located closer toward the GC  will first fall onto
black hole rather than evaporate. 

Thermal instability mechanism explains in the natural way the gas pressure equilibrium 
between the hot and the cold plasma. It also allows to estimate the typical 
size of the cold clump to be of the order of 10$^{14-15}$ cm,
and the total mass of about 10 Earth masses. It will change 
with the shape and strength of radiation field.
But it agrees with recent results by  
\citet{shcherbakov2014}, who  suggested mass for G2 cloud from 4 to 20 M$_{\oplus}$.

We computed most of the results not taking dust into consideration but we showed
that the presence of the dust does not change our main conclusions. The temperature of 
the cold clumps is not affected by dust, since the atomic cooling is also 
very efficient
at temperatures $\sim 10^4$ K. The presence of the dust is extremely important
for the hot material, but the dust embedded in a hot plasma is collisionally 
destroyed in a timescale of a year and after that we only expect the 
coexistence of the cold dusty phase and a hot dustless plasma.  

\section*{Acknowledgments}

This research was supported by  Polish National Science 
Centre grants No. 2011/03/B/ST9/03281, 2013/10/M/ST9/00729, and by Ministry of 
Science and Higher Education grant W30/7.PR/2013.  It received funding
from the European Union Seventh Framework Program (FP7/2007-2013) under 
grant agreement No.312789.  D.K. and V.K.
acknowledge support from the collaboration project between the Czech Science
Foundation and Deutsche Forschungsgemeinschaft (GACR-DFG 13-00070J).

\bibliographystyle{mn2e}
\bibliography{refs}

\end{document}